\documentclass[preprint,showpacs,preprintnumbers,superscriptaddress,amsmath,amssymb]{revtex4}
\usepackage{amssymb}
\usepackage[dvips]{graphicx}% Include figure files
\usepackage{dcolumn}% Align table columns on decimal point
\usepackage{bm}% bold math
\usepackage{subfigure}
\usepackage{floatflt}
\usepackage{float}
\usepackage{rotating}
\usepackage{multirow}
\usepackage[bookmarksnumbered,bookmarksopen,colorlinks,citecolor=blue,linkcolor=blue]{hyperref} %

\begin{document}

\preprint{00-000}

\title{Mass parameters for relative and neck collective motions
        in heavy ion fusion reactions}

\author{Kai Zhao}
\affiliation{China Institute of Atomic Energy, Beijing 102413, P.
R. China}
\author{Zhuxia Li}
\email{lizwux@.ciae.ac.cn} \affiliation{China Institute of Atomic
Energy, Beijing 102413, P. R. China}
\author{Xizhen Wu}
\email{lizwux9@ciae.ac.cn} \affiliation{China Institute of Atomic
Energy, Beijing 102413, P. R. China}
\author{Zhixiang Zhao}
\affiliation{China Institute of Atomic Energy, Beijing 102413, P.
R. China}

\date{\today}% It is always \today, today,

\begin{abstract}
Mass parameters for the relative and neck motions in fusion
reactions of symmetric systems $^{90}$Zr+$^{90}$Zr,
$^{110}$Pd+$^{110}$Pd, and $^{138}$Ba+$^{138}$Ba are studied by
means of a microscopic transport model. The shape of the nuclear
system is determined by an equi-density surface obtained from the
density distribution of the system. The relative and neck motions
are then studied and the mass parameters for these two motions are
deduced. The mass parameter for the relative motion is around the
reduced mass when the reaction partners are at the separated
configuration and increases with decrease of the distance between
two reaction partners after the touching configuration. The mass
parameter for the neck motion first decreases slightly up to the
touching configuration and then increases with the neck width, and
its magnitude is from less than tenth to several times more than
the total mass of the system. The mass parameters obtained from
the microscopic transport model are larger than the ones obtained
from the hydrodynamic model and smaller than those obtained from
the linear response function theory.  The mass parameters for both
motions depend on the reaction systems, but the one for the
relative motion depends on the incompressibility of the EoS more
obviously than that for neck motion.
\end{abstract}

 %25.40.-h Nucleon-induced reactions
 %21.30.-x nuclear forces
 %24.10.Lx Monte Carlo simulations
\maketitle

\section{Introduction}
In the study of a large amplitude collective motion in nuclear
systems including heavy ion fusion reactions( especially the
synthesis of superheavy nuclei) and fission, the macroscopic model
plays a very important role, in which a few collective degrees of
freedom are usually used to describe the complex dynamic process.
The potential energy surface, the mass parameter and the viscosity
are the most important quantities in the macroscopically
description of the large amplitude collective motion. Many works
have been done for the study of these three quantities. The
importance of the potential energy surface in the nuclear large
amplitude collective motion such as in fission and heavy ion
fusion process is well known and up to now the calculation of
potential energy surface based on the macroscopic and microscopic
method seems to be quite successful
\cite{Str67,Mye66,Bra72,Mol01}. But the problem of kinetic energy
,i.e. the mass parameter, is far less settled than that of
potential energies. As a matter of fact, the dynamical behavior of
nuclear systems with large-amplitude collective motions, such as
fusion and fission not only depends on the potential energy
surface but also on mass parameter.  Take a most simple example,
let us consider the fusion process as a barrier penetration
problem. In the WKB formula for barrier penetration the inertia
tensor M(R) appears in the action integral
S(L)=$\mp\frac{2}{\hbar}\int_{L}(2M(R)(U(R)-E))^{1/2}dR$ along the
fusion path L. The U(R) and E are the potential energy surface and
the center of mass energy, respectively. The sign '-' and '+'
refer to above and below the barrier energies. For simplicity only
the internuclear distance R is taken to be the collective
coordinate. The probability for penetrating the barrier(may be a
multi-dimension barrier) is calculated as usual by
P(L)=exp(-S(L)). In many studies the M(R) was simply taken to be
the reduced mass. This simple approximation is only partly valid
for fusion reaction of light nuclei. For fusion reactions of heavy
nuclei, it is well known that not only the relative collective
motion but also the neck motion becomes very important. The
process for penetrating the barrier becomes more complicated and
the M(R) can not be simply taken to be the reduced mass. Thus, the
impact of the mass parameter on fusion process for heavy systems
becomes incontestable. Recently renewed interest in the study of
mass parameter is motivated by the study of the synthesis of
superheavy nuclei\cite{Ada99,Ada00,Zag01,She02}. The qualitative
and quantitative study of the mass parameter becomes more
important. In the dynamic study of synthesis of superheavy nuclei
the mass parameters for collective motions were studied by the
cranking model( or by the linear response function theory(LRFT)
based on the cranking model) and the hydrodynamical model. In
\cite{Ada99,Ada00}, a large mass parameter for neck motion was
found by means of the LRFT method leading to a restriction for a
growing neck in dinuclear system and melting of the dinuclear
system along the internuclear distance into a compound system.
While in \cite{She02} a hydrodynamical mass parameter was adopted,
then a subsequent shape evolution of pear shaped mono-nucleus
formed with two incident ions could happen. Thus, the rather
different mass parameters obtained by cranking and hydrodynamical
models lead to complete different description of the fusion
mechanism for heavy systems due to different fusion paths. The
cranking model was developed to calculate the mass parameter of
nuclear systems
microscopically\cite{Ing54,Bel59,Tho60,Hof97,Cas85,Fen84}. But the
abrupt change of mass parameters near the level crossing makes
difficulties for practical usage\cite{Sch86}. Whereas for the
hydrodynamical model, the assumption of the irrotational flow of
an incompressible and non-viscous fluid
\cite{Nix67,Wer69,Wu87,She02,Ghe06} is too simple for many
practical purposes. Furthermore, in both approaches, one usually
adopts a way that only one collective degree of freedom is taken
to be a variable and the others are fixed in the calculation of
mass parameters. The condition of the validity of this (static)
treatment needs to be tested because in a real heavy ion fusion
reaction all degrees of freedom of collective motion change
dynamically and self-consistently. Therefore, it seems to be
necessary to investigate the mass parameters for the collective
motion in fusion reactions by a more realistic dynamic model. In
this work for the first time we try to employ the microscopic
transport model, namely the Improved Quantum Molecular Dynamics
(ImQMD) model to investigate the mass parameters for collective
motions in heavy ion fusion reactions and possibly to find out the
conditions for the validity of the widely used static treatment in
the calculations of mass parameter.

The quantum molecular dynamics (QMD) model being successfully used
in intermediate energy heavy-ion collisions was extended to apply
to heavy ion collisions at energies near barrier by making a
series of improvements\cite{Wan02,Wan04}. The main improvements
introduced are: introducing the surface and surface symmetry
energy terms in the potential energy density functional,
introducing the phase space occupation constraint\cite{Pap01}
which can be considered as an approximate treatment of
anti-symmetrization, and a system size dependent wave packet
width. Furthermore, considering the especially importance of a
proper initial condition for the application of the microscopic
transport model to low energy heavy ion reactions, in the ImQMD
model the requirement of the projectile and target at initial time
being at ground states is enforced. And consequently the
projectile and target can be stable for a long enough time and the
structure effect in entrance can be taken into account partly.
Thus, in principle, the dissipation, diffusion and correlation
effects are all included without introducing any freely adjusting
parameter. With the ImQMD model the fusion dynamics at energies
near and above the barrier has extensively been studied. Firstly,
the stable initial nuclei with good ground-state properties such
as the binding energies, root-mean-square radii and density
distribution for a series of nuclei from $^{16}$O to $^{208}$Pb
and $^{238}$U were realized within several thousands fm/$c$
\cite{Wan02,Wan05}. The Coulomb barriers for many reaction
partners can be described well. The model were applied to
calculate fusion (capture) excitation functions for a series of
fusion reactions including neutron-rich projectile and target
\cite{Wan04}. We also extend the ImQMD model to study the capture
cross sections for the reactions $^{48}$Ca+$^{208}$Pb,$^{238}$U
leading to super-heavy nuclei \cite{Tian08}. The calculated
results are in good agreement with experimental data. Thus, it is
more confident to apply the ImQMD model to study the mass
parameters for collective motions in heavy ion fusion reactions.
In this work, the mass parameters for symmetric reactions of
$^{90}$Zr+$^{90}$Zr, $^{110}$Pd+$^{110}$Pd, and
$^{138}$Ba+$^{138}$Ba are investigated. The paper includes
following parts: In section II we briefly introduce the ImQMD
model. In section III, we present the results of our study on the
mass parameters for $^{90}$Zr+$^{90}$Zr, $^{110}$Pd+$^{110}$Pd,
and $^{138}$Ba+$^{138}$Ba. Finally, we give brief summary and
discussion in section IV.
\section{Brief introduction of the ImQMD model}
In the ImQMD model as the same as the original QMD model, each
nucleon is represented by a coherent state of a Gaussian wave
packet \cite{Hart89,Ai91},
\begin{equation}  \label{1}
\phi _{i}(\mathbf{r})=\frac{1}{(2\pi \sigma _{r}^{2})^{3/4}}\exp [-\frac{(%
\mathbf{r-r}_{i})^{2}}{4\sigma
_{r}^{2}}+\frac{i}{\hbar}\mathbf{r}\cdot \mathbf{p}_{i}],
\end{equation}
where, $\mathbf{r}_{i}, \mathbf{p}_{i}$, are the center of i-th
wave packet in the coordinate and momentum space, respectively.
$\sigma _{r}$ represents the spatial spread of the wave packet.
 Through a Wigner transformation, the one-body
 phase space distribution function for N-distinguishable particles is
given by:
\begin{equation}  \label{2}
f(\mathbf{r,p})=\sum\limits_{i}\frac{1}{(\pi\hbar)^{3}}\exp[-\frac{(\mathbf{%
r-r}_{i})^{2}}{2\sigma_{r}^{2}}-\frac{2\sigma_{r}^{2}}{\hbar^{2}}(\mathbf{p-p%
}_{i})^{2}].
\end{equation}

The density and momentum distribution functions of a system read
\begin{equation}
\rho (\mathbf{r})=\int f(\mathbf{r,p})d\mathbf{p}=\sum\limits_{i}\rho _{i}(%
\mathbf{r}),  \label{3}
\end{equation}

\begin{equation}
g(\mathbf{p})=\int
f(\mathbf{r,p})d\mathbf{r}=\sum\limits_{i}g_{i}(\mathbf{p}),
\label{4}
\end{equation}
respectively, where the sum runs over all particles in the system.
$\rho _{i}(\mathbf{r})$ and $g_{i}(\mathbf{p})$ are the density
and momentum distributions of nucleon i:
\begin{equation}
\rho _{i}(\mathbf{r})=\frac{1}{(2\pi \sigma _{r}^{2})^{3/2}}\exp
[-\frac{( \mathbf{r-r}_{i})^{2}}{2\sigma _{r}^{2}}],  \label{5}
\end{equation}

\begin{equation}  \label{6}
g_{i}(\mathbf{p})=\frac{1}{(2\pi \sigma _{p}^{2})^{3/2}}\exp [-\frac{(%
\mathbf{p-p}_{i})^{2}}{2\sigma _{p}^{2}}],
\end{equation}
where $\sigma_{r}$ and $\sigma_{p}$ are the widths of wave packets
in coordinate and momentum space, respectively, and they satisfy
the minimum uncertainty relation:
\begin{equation}  \label{7}
\sigma_{r}\sigma_{p}=\frac{\hbar}{2}.
\end{equation}

The propagation of nucleons under the self-consistently generated
mean field is governed by Hamiltonian equations of motion:
\begin{equation}  \label{8}
\dot{\mathbf{r}}_{i}=\frac{\partial H}{\partial \mathbf{p}_{i}}, \dot{%
\mathbf{p}}_{i}=-\frac{\partial H}{\partial \mathbf{r}_{i}}.
\end{equation}
Hamiltonian $H$  consists of the kinetic energy and effective
interaction potential energy,
\begin{equation}  \label{9}
H=T+U_{loc}+U_{Coul}.
\end{equation}
%\begin{equation}  \label{10}
%T=\sum\limits_{i} \frac{\mathbf{p}_{i}^{2}}{2m}.
%\end{equation}
%The effective interaction potential energy includes the nuclear
%local interaction potential energy and Coulomb interaction
%potential energy,
%\begin{equation}
%U=U_{loc}+U_{Coul}.  \label{11}
%\end{equation}%
%And
%\begin{equation}
%U_{loc}=\int V_{loc}(\mathbf{r})d\mathbf{r},  \label{12}
%\end{equation}
Here, $U_{Coul}$ is the Coulomb energy,
    and ${\it{U_{loc}}}=\int V_{loc}[\rho(\bm{r})] d \bm{r}$. $V_{loc}[\rho(\bm{r})]$ is the nuclear
    potential energy density functional, which is introduced according to Skyrme interaction energy density
functional with spin-orbit term omitted \cite{Wan04,Zha06}. It
reads
\begin{equation}
V_{loc}[\rho]=\frac{\alpha }{2}\frac{\rho ^{2}}{\rho _{0}}+\frac{\beta }{\gamma +1}%
\frac{\rho ^{\gamma +1}}{\rho _{0}^{\gamma }}+\frac{g_{sur}}{2\rho _{0}}%
(\nabla \rho )^{2}+\frac{%
C_{s}}{2\rho _{0}}(\rho ^{2}-\kappa _{s}(\nabla \rho )^{2})\delta
^{2}+g_{\tau }\frac{\rho ^{\eta +1}}{\rho _{0}^{\eta }}.
\label{13}
\end{equation}%
Here, $\delta=(\rho_{n}-\rho_{p})/(\rho_{n}+\rho_{p})$ is the
isospin asymmetry and the $\rho$, $\rho_{n}$, $\rho_{p}$ are the
nucleon, neutron, and proton density, respectively. By integrating
$V_{loc}$, we obtain the local interaction potential energy:

\begin{eqnarray}
U_{loc} &=&\frac{\alpha }{2}\sum\limits_{i}\sum\limits_{j\neq
i}\frac{\rho _{ij}}{\rho _{0}}+\frac{\beta }{\gamma
+1}\sum\limits_{i}\left( \sum\limits_{j\neq i}\frac{\rho
_{ij}}{\rho _{0}}\right) ^{\gamma }
\nonumber \\
&&+\frac{g_{sur}}{2}\sum\limits_{i}\sum\limits_{j\neq i}f_{sij}\frac{\rho _{ij}}{%
\rho _{0}}+g_{\tau }\sum\limits_{i}\left( \sum\limits_{j\neq
i}\frac{\rho
_{ij}}{\rho _{0}}\right) ^{\eta }+\frac{C_{s}}{2}\sum\limits_{i}\sum%
\limits_{j\neq i}t_{i}t_{j}\frac{\rho _{ij}}{\rho _{0}}\left(
1-\kappa _{s}f_{sij}\right),  \label{14}
\end{eqnarray}%

where

\begin{equation}
\rho_{ij}=\frac{1}{(4\pi\sigma_{r}^{2})^{3/2}}exp[-\frac{(\mathbf{r}_{i}-\mathbf{r}_{j})^{2}}{4\sigma_{r}^{2}}],\label{15}
\end{equation}

\begin{equation}
f_{sij}=\frac{3}{2\sigma_{r}^{2}}-(\frac{\textbf{r}_{i}-\textbf{r}_{j}}{2\sigma_{r}^{2}})^{2},\label{16}
\end{equation}
and $t_{i}$=1 and -1 for proton and neutron, respectively.

The Coulomb energy can be written as the sum of the direct and the
exchange contribution
\begin{equation}
U_{Coul}=\frac{1}{2}\int \int \rho _{p}(\mathbf{r})\frac{e^{2}}{|\mathbf{r-r}%
^{\prime }|}\rho _{p}(\mathbf{r}^{\prime
})d\mathbf{r}d\mathbf{r}^{\prime }-e^{2}\frac{3}{4}\left(
\frac{3}{\pi }\right) ^{1/3}\int \rho _{p}^{4/3}d\mathbf{r}.
\label{17}
\end{equation}%
 The collision term and the constraint for single particle occupation
number are considered as the same as that in
\cite{Wan02,Wan04,Pap01}.

To have a proper initial condition ( with good properties of
projectile and target nuclei) is of crucial importance for
studying low energy heavy ion reactions by means of the transport
model description. In this work we pay special attention to the
initial condition.  The procedure for making initial nuclei is
similar as that in ref.~\cite{Wan02}. The initial nuclei applied
in the study of reaction process have good ground state properties
such as binding energies and rms radii and furthermore, their time
evolution is very stable remaining almost unchanged for a long
enough time(in this work,it is taken to be 2000fm/c). At the same
time they are required to be without spurious particle emission.
For the self-consistency, we adopt the same effective nuclear
potential energy density in making initial nuclei and the
simulation of the reaction process when solving the Hamiltonian
equation(8). Two sets of interaction force parameters used in the
calculations and the corresponding properties of saturated nuclear
matter are given in table \ref{parameter}. In this work, the
definition of a fusion event is taken to be the same as that in
\cite{Wan02}, in which a fusion event is defined operationally
like in TDHF calculations that is that the fusion event is defined
rather operationally as the event in which the coalesced one-body
density survives through one or more rotations of composite system
or through several oscillations of its radius.

\begin{table}[htbp]
\vspace{10mm}
\begin{center}
\begin{tabular}{ccccccccccc}
 \hline
  & $\alpha$(MeV) & $\beta$(MeV) & $\gamma$ & $g_{sur}$(MeV$fm^{2}$) & $g_{\tau}$(MeV) & $\eta$ & $c_{s}$(MeV) & $\kappa_{s}$($fm^{2}$) & $\rho_0$($fm^{-3}$) & $K_{\infty}$($MeV$)\\
  \hline
  IQ1 & -310 & 258 & 7/6 & 19.8 & 9.5 & 2/3 & 32 & 0.08 & 0.165 & 144\\
  \hline
  IQ2 & -356 & 303 & 7/6 & 7.0 & 12.5 & 2/3 & 32 & 0.08 & 0.165 & 175\\
  \hline
\end{tabular}
\end{center}
\caption{\label{parameter}the model parameters}
\end{table}
\section{the mass parameters for collective motions in heavy ion fusion reactions}
In the macroscopic model, the low energy heavy ion fusion
reactions are generally described by five collective coordinates,
namely the distance between the centers of mass of two colliding
nuclei, the mass asymmetry degree of freedom, the neck parameter
and deformation coordinates of two ends of the colliding system.
Since in the early stage of reactions the mass parameters for
relative collective motion and neck motion are weakly dependent on
the deformations of two end parts\cite{Ada00}, one often considers
only three collective modes, i.e., the relative motion of two
nuclei, the formation and rupture of neck and nucleon transfer
between projectile and target. The mass parameters for those
collective motion modes are expressed by a matrix and the
corresponding collective kinetic energy is given by

\begin{equation}
T=\frac{1}{2}\sum_{ij}M_{ij}(q)\dot{q}_i\dot{q}_j.
\end{equation}
Here, $M_{ij}$ is the mass parameter matrix and $q_{i}$ is the
i-th collective coordinate. The nuclear mass parameters reflect
the properties of a nuclear system related to the collective
kinetic motion, which depend on the time dependent change of the
shape of nuclear system and may depend on the effective
nucleon-nucleon interaction. For the symmetric reactions
considered in this work the mass asymmetric motion plays a
negligible role and thus we can only consider the relative motion
of projectile and target and the neck motion in the following
study.
\subsection{The mass parameter for the relative motion of projectile(like) and
target(like)} 
\begin{figure}[h]
    \centering
        \includegraphics[angle=270,width=0.5\textwidth]{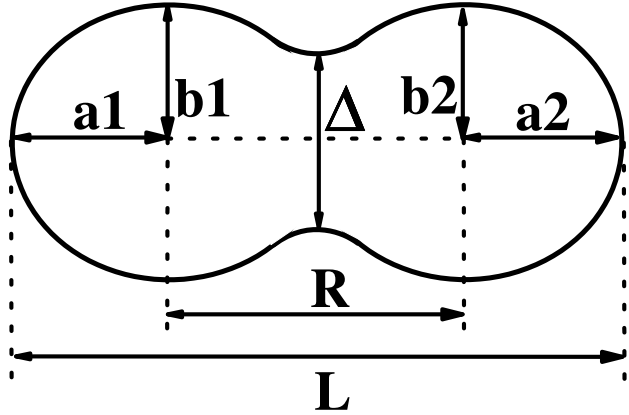}
        \caption{(Color online) The schematic figure of deformation parameters of a reaction
system.}
        \label{fig1}
    \end{figure}

Let us first investigate the mass parameter for
relative motion of projectile and target nuclei. In order to study
the collective motion with the microscopic transport model one has
to first define the collective coordinate by means of the
microscopic quantities. The distance between the centers of mass
of two nuclei $R$ can be expressed as
\begin{equation}
R=A^{-1}\int{z\cdot{sign(z)}f(\vec{r},\vec{p})d\vec{r}d\vec{p}}.
\end{equation}
Here, $sign(z)=\left\{%
\begin{array}{ll}
    +1, & \hbox{z$>$0;} \\
    -1, & \hbox{z$<$0.} \\
\end{array}%
\right. $ and the $f(\vec{r},\vec{p})$ is the distribution
function in phase space. The distance between the centers of mass
of two nuclei $R$ and the neck width $\Delta$ (which will be
discussed in the next subsection) for an axial symmetric nuclear
system are sketched in Fig.\ref{fig1}. In addition, the system
length L is also shown in Fig.\ref{fig1}. Within the ImQMD model
approach, the conjugate momentum of the relative motion of two
nuclei is expressed as
\begin{equation}
P_{R}=\int{P_{z}\cdot{sign(z)}f(\vec{r},\vec{p})d\vec{r}d\vec{p}}.
\end{equation}

 The velocity of the relative motion of two nuclei
\begin{equation}
\dot{R}=\frac{\Delta{R}}{\Delta{t}}
\end{equation}
can be obtained numerically. The mass parameter for relative
collection motion of two nuclei is then calculated by
\begin{equation}
M_{RR}=\frac{P_{R}}{\dot{R}}.
\end{equation}
This formula seems to be questionable when the $\dot{R}$
approaches to zero, but , in fact, the $P_{R}$ also approaches to
zero as the $\dot{R}$ $\rightarrow$ 0. This situation only happens
at the fusion reaction closing to the end, which is a not
interesting case.

 In this work the fusion systems of
$^{90}$Zr+$^{90}$Zr, $^{110}$Pd+$^{110}$Pd, $^{138}$Ba+$^{138}$Ba
are chosen for studying mass parameters. They are all along the
$\beta$-stability line. The head on collisions at the incident
energy of 1.1 times the height of Bass barrier\cite{Bass} are
considered for all systems. The incident energy is selected in
such a way that the fusion process can be proceeded, whereas the
compress or expand of nuclear matter does not appear.

    \begin{figure}[h]
    \centering
        \includegraphics[angle=270,width=0.6\textwidth]{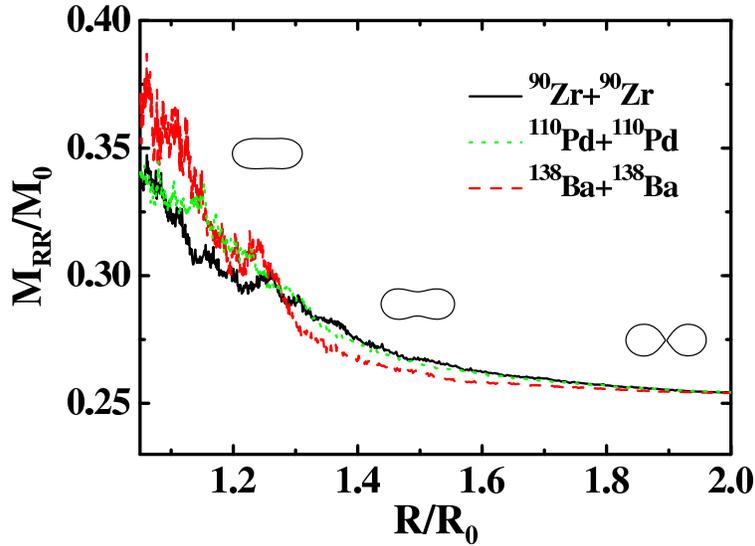}
        \caption{(Color online)The mass parameters for the relative
    collection motion as a function of $R/R_{0}$ for systems of
    $^{90}$Zr+$^{90}$Zr, $^{110}$Pd+$^{110}$Pd and $^{138}$Ba+$^{138}$Ba.}
        \label{fig2}
    \end{figure}

Fig.\ref{fig2} shows the dependence of the mass parameter $M_{RR}$
in the unit of total mass of the reaction system $M_{0}$ on the
scaled distance between the centers of mass of two nuclei
$R/R_{0}$ for systems of $^{90}$Zr+$^{90}$Zr,
$^{110}$Pd+$^{110}$Pd and $^{138}$Ba+$^{138}$Ba calculated with
IQ2. Here $R_{0}$= $r_{0}(A_{1}+A_{2})^{1/3}$fm is the radius of
the spherical compound nucleus formed by two nuclei. The $r_{0}$
is taken to be 1.16 fm. The corresponding configurations of the
reaction system at different stages are also shown in the figure.
The structures appeared in the $M_{RR}$ in the figure, especially
for $^{138}$Ba+$^{138}$Ba is mainly due to not enough number of
the fusion events. From the figure one sees that when $R
> 1.8R_{0}$ the mass parameter $M_{RR}$ approaches to 0.25 total
mass of the system, i. e., the reduced mass, whereas in the case
of $R < 1.8R_{0}$ the $M_{RR}$ increases with decrease of the
distance R. Another important feature is that the mass parameter
for relative motion of two nuclei depends on the size of the
reaction systems. The heavier system has the steeper increasing
slope of $M_{RR}$ with decrease of $R$.

The incident energy dependence of mass parameters is also
interesting. Here we only study the energy dependence of the
$M_{RR}$. Fig.3 shows the mass parameter for relative motion for
$^{110}$Pd+$^{110}$Pd at E=1.0, 1.1 and 1.5V$_{B}$, where V$_{B}$
refers to the height of the Bass barrier. The figure shows that
the mass parameter for relative motion depends on the incident
energy. For lower energy like E=1.0V$_{B}$, the mass parameter for
relative motion changes slowly since the reaction process is slow
and the reaction system has time to change their shapes gradually
and consequently the mass parameter for relative motion increases
gradually from reduced mass to larger than the reduced mass. As
the incident energy increases the reaction process becomes faster
and there is less time to change the shape of the system and the
mass parameter for relative collective can be close to the reduced
mass for a period of time and then increases rapidly at very late
time. So we see from the figure that at higher energy such as at
E=1.5V$_{B}$ the $M_{RR}$ is roughly equal to reduced mass when
R/R$_{0}\geq1.3$. But we expect it could be the case only for not
heavy nuclear systems and at very above the barrier energies. The
time evolution of the mass parameters for relative motion at three
energies is shown in the inserted figure of Fig.3.
\begin{figure}[h]
    \centering
        \includegraphics[angle=270,width=0.6\textwidth]{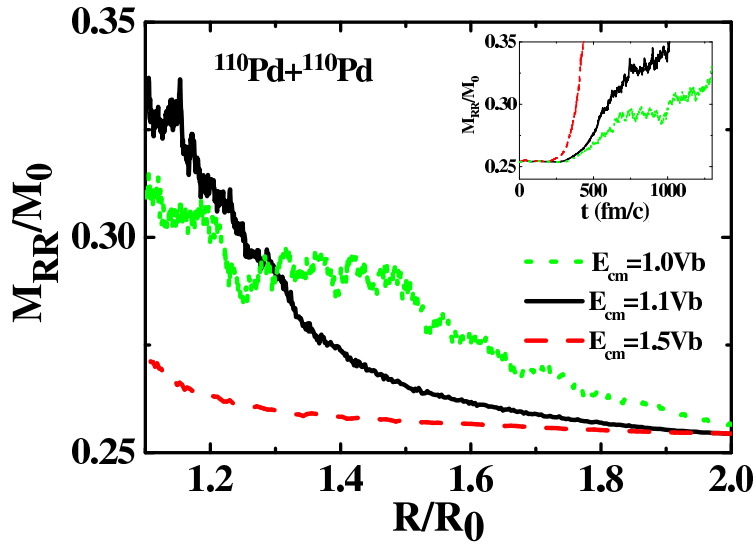}
        \caption{(Color online)The incident energy dependence of the $M_{RR}$ for
$^{110}$Pd+$^{110}$Pd at E=1.0,1.1,1.5V$_{B}$. The inserted figure
is the time evolution of the $M_{RR}$ for the same system.}
        \label{fig3}
    \end{figure}

\begin{figure}[h]
    \centering
        \includegraphics[angle=270,width=0.6\textwidth]{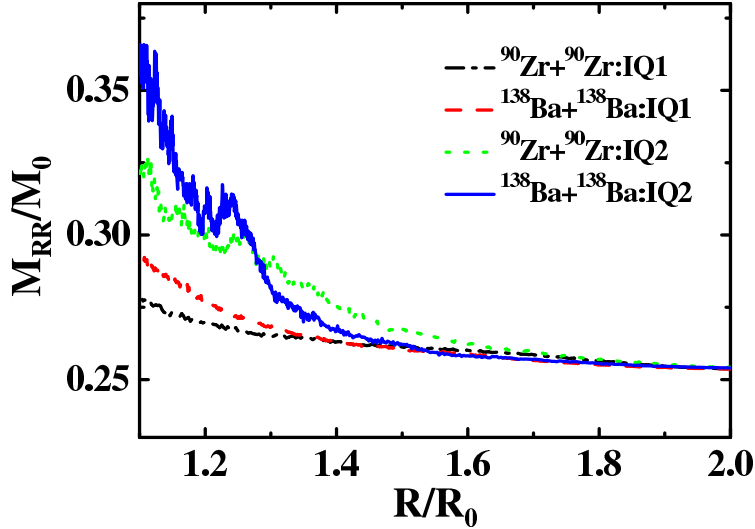}
        \caption{(Color online)The comparison between mass parameters $M_{RR}$
    calculated with force parameter sets IQ1 and IQ2.}
        \label{fig4}
    \end{figure}
Now let us investigate whether the mass parameter for relative
motion depends on the nucleon-nucleon effective interaction. In
Fig.\ref{fig4} we make a comparison between the mass parameters
$M_{RR}$ calculated with two force parameter sets IQ1 and IQ2
(listed in Table \ref{parameter}) for reaction systems of
$^{90}Zr+^{90}Zr$ and $^{138}Ba+^{138}Ba$. One sees that the
$M_{RR}$ calculated with IQ1 rises slower than that calculated
with IQ2. We notice that the parameter set IQ1 gives relatively
smaller incompressibility compared with the force IQ2 (see Table
\ref{parameter}). The difference between the $M_{RR}$ calculated
with IQ1 and IQ2 can be attributed to the difference in
incompressibility of interaction parameters IQ1 and IQ2. The
softer EoS needs less energy to make two nuclei even closer than
touching configuration and thus the mass parameter for relative
motion for the fusion reaction system decreases with the decrease
of the stiffness of the EoS.

    \begin{figure}[h]
    \centering
        \includegraphics[angle=270,width=0.6\textwidth]{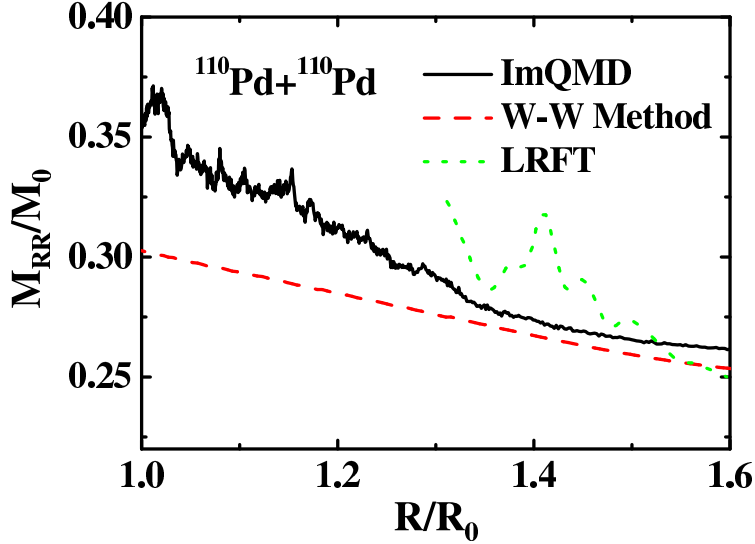}
        \caption{(Color online)The comparison of mass parameter $M_{RR}$
       calculated by the ImMQD model with that by Werner-Wheeler method
       and the linear response function theory.}
        \label{fig5}
    \end{figure}

In order to make comparison with other approaches, in
Fig.\ref{fig5} we show the results calculated with hydrodynamic
model(Werner-Wheeler (W-W) method) \cite{Wu87,She02} and with the
linear response function theory model(LRFT)\cite{Ada00}, in
addition to the microscopic transport model approach for the
system of $^{110}$Pd+$^{110}$Pd. The results obtained by the ImQMD
model with IQ2 and by hydrodynamic model(W-W) are shown by solid
and long dashed curves, respectively. The short dashed curve is
the results by LRFT taken from \cite{Ada00} where the large
fluctuation appears due to the level crossing. The general
tendency of the $M_{RR}$ changing with R/R$_{0}$ calculated by the
ImQMD model is similar with those obtained by hydrodynamic model
and by LRFT. The magnitude of the mass parameter obtained by the
ImQMD model is higher than that obtained by hydrodynamic model,
but lower than that obtained by LRFT.

\subsection{The mass parameter for the neck motion in fusion reactions}
\begin{figure}[h]
    \centering
        \includegraphics[angle=270,width=0.6\textwidth]{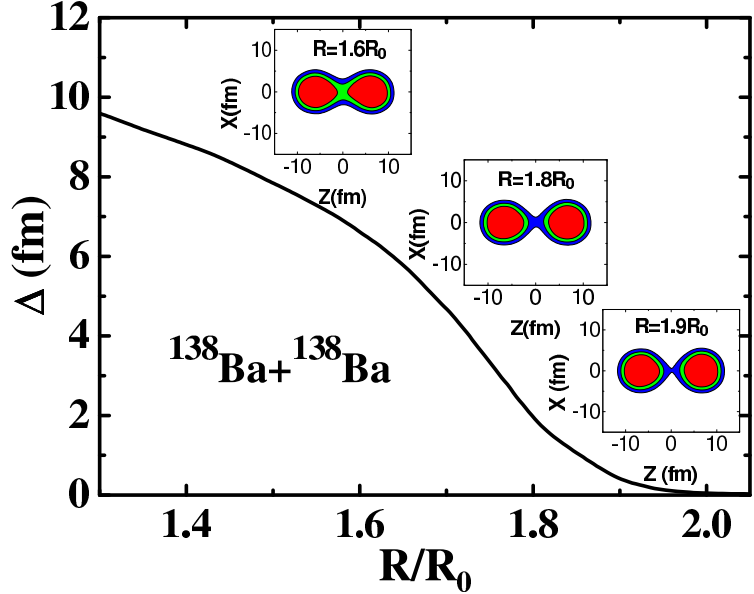}
        \caption{(Color online)The correlation between neck width $\Delta$ and
     distance R/R$_{0}$ for $^{138}Ba+^{138}Ba$.}
        \label{fig6}
    \end{figure}

 In order to study the mass parameter for the neck motion within microscopic transport model one has
 to first define a nuclear surface based
on the density distribution of the reaction system. The shape of
the nuclear system is then defined by the nuclear surface. The
nuclear surface is usually defined by an equi-density surface of
the reaction system. In this work this equi-density surface,
namely the $\rho_{s}(z,t)$ for a axial symmetric system at time t
is taken to be at the half normal density $\rho=0.5\rho_{0}$. The
neck width $\Delta$ is defined by 2$\rho_{s}(z=z_{c},t)$ and the
$z_{c}$ denotes the position of the neck as shown in
Fig.\ref{fig1}.
 Thus, the distance of two nuclei and neck width are evolved
with time simultaneously. Fig.\ref{fig6} shows the correlation
between the neck width $\Delta$ and distance between two nuclei R/
R$_{0}$. The contour plots of density distribution for
$^{138}$Ba+$^{138}$Ba at R/ R$_{0}$=1.9, 1.8, 1.6 are also shown
in the figure, the lines in the contour plots correspond to
$\rho$=0.5, 0.75, 1.0 $\rho_{0}$, respectively. One can find that
the touching configuration is around R/ R$_{0}$=1.8, but not at R/
R$_{0}$=1.6 which is about the sum of the radii of two nuclei.
This is because two nuclei are elongated due to the interaction
between two nuclei. One can further find that the most fast
increase of the neck width $\Delta$ happens around the touching
configuration, for instance, in the system $^{138}Ba+^{138}Ba$ the
$\Delta$ from 0.5fm increases to about 7fm when the R/ R$_{0}$
decreases only from 1.9 to 1.6. It means that the touching
configuration is important for characterizing the neck motion.

\begin{figure}[h]
    \centering
        \includegraphics[angle=270,width=0.6\textwidth]{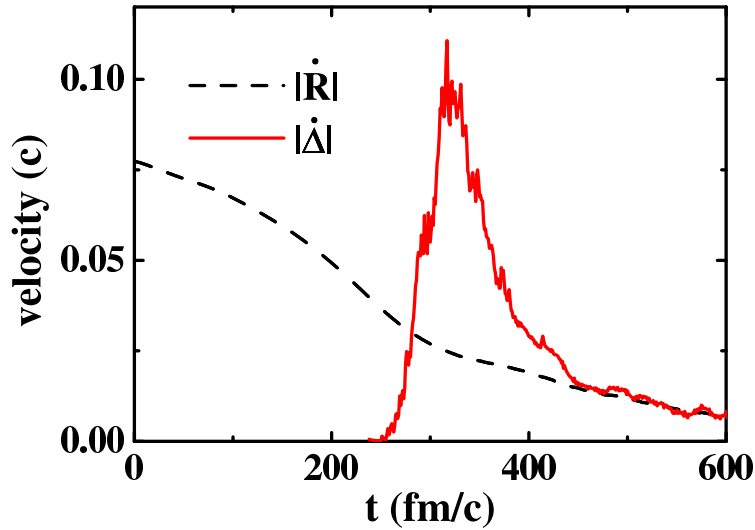}
        \caption{(Color online)The time evolution of the $\dot{\Delta}$ and
     $\dot{R}$ for $^{138}$Ba+$^{138}$Ba. The solid
     curve is for $\dot{\Delta}$ and the dashed curve for $\dot{R}$.}
        \label{fig7}
    \end{figure}

Mass parameters, in fact, are related with the kinetic energies of
collective motions. The collective kinetic energy for relative
motion of two colliding nuclei and neck motion can be expressed by
\begin{equation}
T=T_{RR}+T_{R\Delta}+T_{\Delta\Delta}=\frac{1}{2}[M_{RR}\dot{R}^2+2M_{R\Delta}\dot{R}\dot{\Delta}+M_{\Delta\Delta}\dot{\Delta}^2].
\end{equation}
As is indicated in ref.\cite{Sch86} and \cite{Ada99} that for the
mass symmetric reactions the mixing mass parameter involving the
neck degree of freedom remains comparatively small. It means that
the condition $M_{R\Delta}/$ $\sqrt{M_{RR}M_{\Delta\Delta}} << 1$
is approximately satisfied\cite{Ada99}. In order to test whether
the crossing term in (20) can be eliminated or not, in
Fig.\ref{fig7} we show the time evolution of $\dot{R}$ and
$\dot{\Delta}$ for $^{138}$Ba+$^{138}$Ba. It is seen from the
figure that the $\dot{\Delta}$ is a strongly peaked function of
time, and the peak is at about 350 fm/c with very steep rising
left side, while $\dot{R}$ is a monotonically decreasing function
of time. The $\dot{R}$ reduces considerably at 280 fm/c, which is
much earlier than the time when the $\dot{\Delta}$ reaches the
peak value. From this figure we find that the coupling between the
velocities of the relative motion and neck motion is indeed not
strong within the time period which we are interested in and thus,
neglecting the crossing term in (20) is reasonable. Here we should
mention that the velocities for neck and relative motion are
strongly correlated when t$>$450fm/c seen from the figure. We
should also notice that the neck is already well developed at that
time and in this case it is difficult to well distinguish the
relative and neck motion and the treatment of neglecting the
crossing term adopted in this work only gives approximate results
.

Now let us investigate the mass parameter for the neck motion in
symmetric fusion systems. Instead of using the relation of
M$_{\Delta\Delta}$=P$_\Delta$/$\dot{\Delta}$ we calculate
M$_{\Delta\Delta}$ through the relation

\begin{equation}
T_{\Delta\Delta}=\frac{1}{2}M_{\Delta\Delta}\dot{\Delta}^2
\end{equation}
and
\begin{equation}
T_{\Delta\Delta}=E_{tot}-E_{pot}-T_{RR}-T_{oth}-E_{exc}.
\end{equation}

\begin{figure}[h]
    \centering
        \includegraphics[angle=270,width=0.8\textwidth]{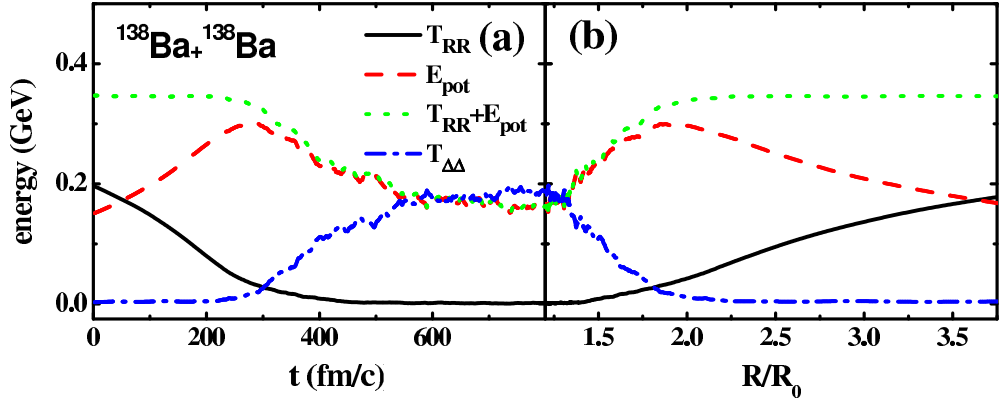}
        \caption{(Color online)The kinetic energies for the relative and neck motions
     and the interaction potential energy for the reaction $^{138}$Ba+$^{138}$Ba at $E_{cm}$=350 MeV (a) as a function
     of time and (b)as a function of R/$R_{0}$ .}
        \label{fig8}
    \end{figure}

Here $T_{oth}$ is the kinetic energy from the motion of the
additional degrees of freedom except the relative motion and neck
motion. Since in the early stage of fusion reaction the change of
the deformations at two ends of colliding system is not strong,
the energy corresponding to this motion could be
neglected\cite{Ada00}. Thus the $T_{oth}$ is mainly from the
crossing term of relative motion and neck motion. As is mentioned
above this term is small, then the $T_{oth}$ should be small and
could be neglected. The $E_{exc}$ is the internal excitation
energy. As we know that the neck motion plays important role only
near touching configuration when the internal excitation is weak,
in addition the incident energy is selected to be low, the
internal excitation energy should be much smaller than the
collective motions. Thus, we neglect the E$_{exc}$ in the present
calculations for simplicity. The $E_{pot}$ is calculated by
\begin{equation}
E_{pot}(R)=E_{12}(R)-E_{1}-E_{2}.  \label{22}
\end{equation}%
Here, $E_{12}(R)$, $E_{1}$ and $E_{2}$ are the total energy of the
whole system, the energies of the projectile(like) and target
(like) part, respectively. The reactions $^{90}$Zr+$^{90}$Zr and
$^{138}$Ba+$^{138}$Ba at incident energies equal to 1.08 times the
height of the Bass barrier(E$_{cm}$=350 MeV) are selected to study
the mass parameter for neck motion. Here the incident energy
chosen is lower than that in the investigation of the mass
parameter for relative collection motion in order to reduce the
internal excitation energy. Fig.\ref{fig8}(a) shows the time
evolution of kinetic energies for relative and neck motions, and
of the potential energy for $^{138}$Ba+$^{138}$Ba at E$_{cm}$=350
MeV. The parameter set IQ2 is utilized. Here the initial time is
taken at the time when the surface to surface distance of two
nuclei equals to 22.5fm (correspond to R/$R_{0}$=3.6 in
Fig.\ref{fig8}(b)). One sees from the figure that the kinetic
energy for relative motion decreases firstly and then approaches
to zero with time increasing, while the kinetic energy for neck
motion is about zero at beginning, then it increases and until 550
fm/c it saturates. The potential energy is zero at the infinite
distance, then it increases as two nuclei approach with each other
and it reaches a maximum value at about 300 fm/c, after then it
reduces to a saturated value. In Fig.\ref{fig8}(b) we show the
kinetic energies for relative motion and neck motion and the
potential energy as a function of R/R$_{0}$. The energy for
relative motion gradually decreases with R/R$_{0}$ decreasing and
finally equals to zero, while the energy for neck motion starts to
increase quickly from zero to a saturated value from R/R$_{0}\sim$
1.9 to R/R$_{0}\sim$ 1.3. We find that the magnitude of the
kinetic energy for neck motion exceeds that of relative motion
just after the touching configuration. The kinetic energies for
both relative and neck motion are saturated at about
R/R$_{0}$=1.3. It is also noticed that the potential energy also
reaches a maximum value near the touching configuration.

\begin{figure}[h]
    \centering
        \includegraphics[angle=270,width=0.8\textwidth]{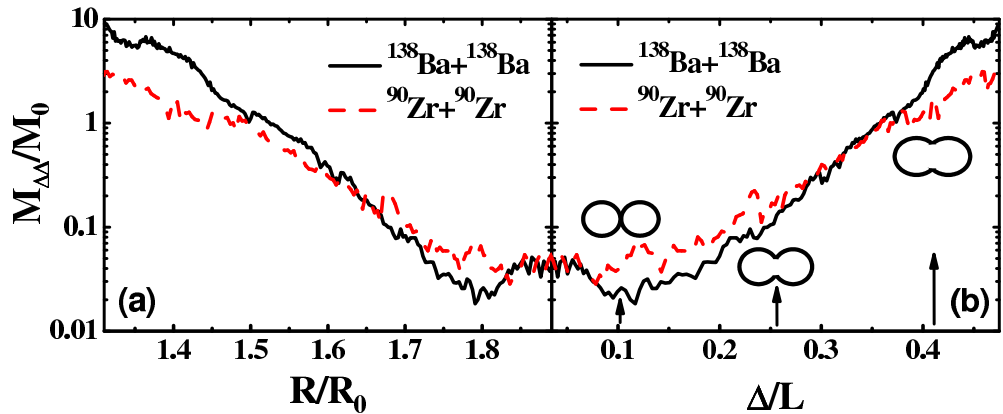}
        \caption{(Color online)The mass parameter M$_{\Delta\Delta}$ for the
     neck motion as a function of a)R/R$_{0}$ and b)$\Delta/L$ for systems
     $^{138}Ba+^{138}Ba$ and $^{90}$Zr+$^{90}$Zr calculated with IQ2.}
        \label{fig9}
    \end{figure}

Having the kinetic energy for neck motion $T_{\Delta\Delta}$ and
velocity $\dot{\Delta}$, we can calculate the mass parameter for
neck motion through expression (21). Fig.\ref{fig9}(a) presents
the evolution of the mass parameter for neck motion with the
relative distance between the centers of mass of two nuclei and
Fig.\ref{fig9}(b) shows the mass parameters as a function of the
neck width scaled by L, the total length of the colliding system,
for the systems $^{90}Zr+^{90}Zr$, $^{138}Ba+^{138}Ba$,
respectively. From Fig.\ref{fig9} one sees that the
$M_{\Delta\Delta}$ reaches a small minimum near the touching
configuration. Then it increases with decrease of R/R$_{0}$ and
increase of $\Delta$. We notice the change in the slope of
$M_{\Delta\Delta}$ with respect to R/R$_{0}$ and $\Delta$ near the
touching configuration, which is due to the sharp peak of
$\dot{\Delta}$ as shown in Fig.7. The magnitude of the
$M_{\Delta\Delta}$ is from less than tenth of the system mass
($A_{1}+A_{2}$) to several times the system mass. Finally it
approaches to saturate values for both systems of
$^{90}Zr+^{90}Zr$ and $^{138}Ba+^{138}Ba$. The figure shows that
slopes for $M_{\Delta\Delta}$ vs $\Delta$(R/R$_{0}$) for
$^{90}Zr+^{90}Zr$ and $^{138}Ba+^{138}Ba$ are different and it
implies the system dependence of the mass parameter for neck
motion. We have also investigated the influence of the effective
interaction on the mass parameter for neck motion by comparing the
mass parameters calculated with IQ1 and IQ2 and we find that there
is no big difference between them. It seems to us that the
dependence of the mass parameter for neck motion on the effective
interaction is weaker than that for the relative collection motion
in nuclear fusion reactions along the $\beta$-stability line.

\begin{figure}[h]
    \centering
        \includegraphics[angle=270,width=0.6\textwidth]{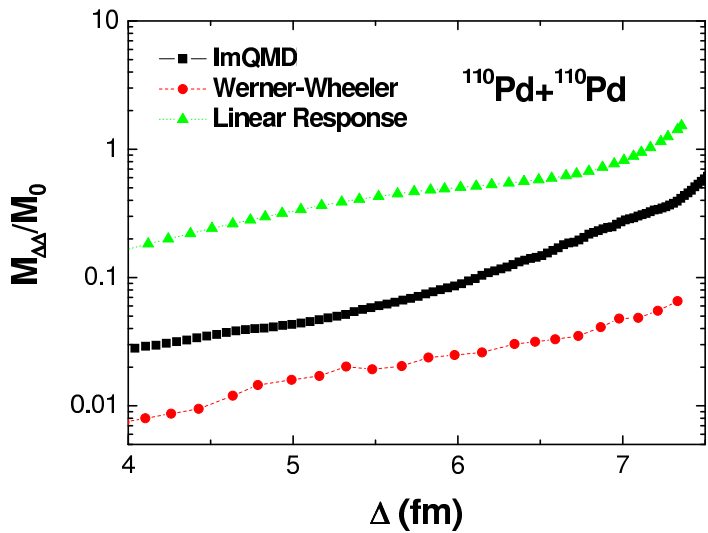}
        \caption{(Color online)The comparison of the mass parameter M$_{\Delta\Delta}$
    calculated by the ImQMD model with that calculated by
    hydrodynamic model and by LRFT taken from \cite{Ada00}.}
        \label{fig10}
    \end{figure}

Finally, we make comparison between the mass parameter for neck
motion obtained by means of the microscopic transport model and
the hydrodynamic model as well as the LRFT. Fig.\ref{fig10} shows
$M_{\Delta\Delta}$/M$_{0}$ as function of neck widths $\Delta$ at
R/R$_{0}$ around the touching configuration. Here, the solid,
dashed, and dotted lines denote the $M_{\Delta\Delta}$ calculated
by the microscopic transport model, the hydrodynamic model and the
results taken from ref. \cite{Ada00} calculated by LRFT,
respectively. The general behavior of the $M_{\Delta\Delta}$ as a
function of neck width is coincident for all results obtained by
three different approaches and also is in agreement with other
studies\cite{Ada99,Sch86}, i.e. the $M_{\Delta\Delta}$ increases
with $\Delta$ and finally it saturates when the neck is already
very much developed. The results of this work are in between the
results from hydrodynamic model and LRFT approaches. This is
similar with the situation in mass parameter for relative motion
as shown in Fig.\ref{fig5}.

\section{summary and discussion}
In summary, in this paper we employ the microscopic transport
model, namely the Improved Quantum Molecular Dynamics model to
study the mass parameters of collective motions in fusion
reactions for the first time. The head on reactions of
$^{90}$Zr+$^{90}$Zr, $^{110}$Pd+$^{110}$Pd, and
$^{138}$Ba+$^{138}$Ba are selected. The essential difference of
the present approach from other methods is that within a
microscopic dynamical model approach, the shape of the reaction
system is determined by the time dependent density distribution of
the system, so that the distance between the centers of mass of
two nuclei and the neck width change with time self-consistently
during the reaction process. It is found that the feature of the
relative and the neck motion is rather different, and the
$\dot{\Delta}$ and the $\dot{R}$ are coupled weakly, as seen from
Fig.\ref{fig7}. So that we can study the $M_{\Delta\Delta}$ and
$M_{RR}$ individually. Then we investigate the mass parameters for
the relative and neck motion for three systems. We find that the
mass parameter for relative motion between two nuclei approaches
the reduced mass when the reaction system is at the separated
configuration and after contact of two partners it increases with
decrease of the distance between two centers of mass. The mass
parameter for neck motion has a small minimum near touching
configuration and after touching configuration it increases with
the increase of neck width(or the decrease of relative distance
between two nuclei). Its magnitude is from less than the tenth to
more than several times the total mass of the system. The general
tendency of the dependence of $M_{RR}$ on the R/$R_{0}$ and
$M_{\Delta\Delta}$ on $\Delta$ is similar with those obtained by
the hydrodynamic model and the LRFT. The magnitude of mass
parameters obtained in this work is larger than the ones obtained
by the hydrodynamic model and smaller than those obtained by LRFT.
Both $M_{RR}$ and $M_{\Delta\Delta}$ depend on the reaction
systems. And the influence of the effective interactions on the
mass parameters is obvious for the mass parameters $M_{RR}$ but
not for $M_{\Delta\Delta}$.

\begin{acknowledgments}
   Authors thank Profs. W. Greiner, W. Scheid and Y.Abe for variable discussions.
   One of authors(Kai Zhao) is grateful to Dr. Caiwan Shen for the help
   in the calculation of the mass parameter within the hydrodynamic model. This
work is supported by the National Natural Science Foundation of
China under Grant Nos.10235030,10675172 and National Basic
Research Programme of China under contract no. 2007CB209900.
\end{acknowledgments}

\end{document}